\documentclass[preprintnumbers,showkeys,superscriptaddress,showpacs,
nofootinbib,byrevtex,fleqn,prd,notitlepage]{revtex4-1}
\usepackage{amsmath,amsfonts,amssymb,amscd,amsxtra,amsthm}
\usepackage{graphicx}
\usepackage{bm,bbold}
\usepackage{epstopdf}
\usepackage{multirow}
\usepackage{hyperref}
\usepackage[fleqn]{nccmath}
\usepackage{empheq}
\usepackage{diagbox}
\input{epsf}
\newcommand{\be}{\begin{eqnarray}}
\newcommand{\ee}{\end{eqnarray}}
\newcommand{\ba}{\begin{array}}
\newcommand{\ea}{\end{array}}

\newcommand{\Slash}[1]{\ooalign{\hfil/\hfil\crcr$#1$}}%%%%TheFeynmanSlash%%%
\usepackage[normalem]{ulem} % \sout{old text} for strikeout
\usepackage[dvipsnames]{xcolor} % For blue in-text comments and% additions
\renewcommand\sout{\bgroup \color{red} \ULdepth=-.5ex \ULset}

\begin{document}
%\preprint{}
\title{On the second Gegenbauer moment of  $\rho$-meson distribution amplitude}
%--------------------------------------------------
\author{Maxim V. Polyakov}
\email[E-mail: ]{maxim.polyakov@rub.de}
\affiliation{Petersburg Nuclear Physics Institute, Gatchina, 188300,
  St. Petersburg, Russia}
\affiliation{Ruhr-Universit\"at Bochum, Fakult\"at f\"ur Physik und Astronomie,
Institut f\"ur Theoretische Physik II, D-44780 Bochum, Germany}
\author{Hyeon-Dong Son}
\email[ E-mail: ]{Hyeon-Dong.Son@ruhr-uni-bochum.de}
\affiliation{Ruhr-Universit\"at Bochum, Fakult\"at f\"ur Physik und Astronomie,
Institut f\"ur Theoretische Physik II, D-44780 Bochum, Germany}
%--------------------------------------------------
\begin{abstract}
Using the soft pion theorem, crossing, and the dispersion relations for the two pion distribution
amplitude ($2\pi$DA) we argue that the  second Gegenbauer moment  the $\rho$-meson DA ($a_2^{(\rho)}$) most probably is negative.
This result is at variance with the majority of the model calculations for  $a_2^{(\rho)}$.

Using the instanton theory of the QCD vacuum, we computed $a_2^{(\rho)}$ at a low normalisation point and obtain for the ratio $ a_2^{(\rho)}/M_3^{(\pi)}$
{\it definitely negative value} in the range of
$a_2^{(\rho)}/M_3^{(\pi)}\in [-2, -1]$.
The range of values corresponds to a generous variation of the parameters of the instanton vacuum.
The value of the second Gegenbauer moment of pion DA is positive in the whole range and is compatible with
its the most advanced lattice measurement.
It seems that the topologically non-trivial field configurations in the
QCD vacuum (instantons) lead to qualitatively different shapes of the pion and the $\rho$-meson DAs.

\end{abstract}
\pacs{}
\keywords{}
\maketitle

\begin{section}{Introduction}
{The meson distribution amplitudes (DAs) describe the lowest Fock component of the meson light-cone wave function.They provide
us with invaluable information about quark-gluon structure of mesons. The meson DAs enter the QCD description of the amplitudes of
various hard exclusive processes, see e. g. the review \cite{Chernyak:1983ej}.   }

The chiral-even leading-twist $\rho$-meson DA $\phi_\rho$ is defined by the following matrix element\footnote{Note that this definition is
not, rigoroursly speaking,  correct: the $\rho$-meson, being a resonance, can not be an asymptotic out state. For a definition of resonance DA see Ref.~\cite{Polyakov:1998ze}.  }
\begin{align}\label{eq:rho_DA_mt}
\left. \langle \rho^0(P) | \;\bar{\psi}(x)\; \Slash{n} \;\frac{\tau^3}{2}\; \psi(0) \;| 0 \rangle \right|_{x^+=x_\perp=0}
= (n \cdot \epsilon^\lambda ) \frac{f_\rho m_\rho}{\sqrt{2}} \int^1_0 dz\; e^{-iz(P \cdot x)} \phi_\rho(z).
\end{align}
Here, $\epsilon^\lambda$ is the polarization vector of the $\rho$-meson and $n$ is a light-like vector.
The dimensional constant $f_\rho$ is defined in a way that the DA follows the normalisation condition:
\begin{align}\label{eq:rho_DA_norm}
\int^1_0 dz\; \phi_\rho(z)=1.
\end{align}
The $\rho$-meson DA can be expanded in terms of Gegenbauer polynomials
\begin{align}\label{eq:rho_DA_G}
\phi_\rho(z)= 6z(1-z)\left(1+\sum^\infty_{n=2,\mathrm{even}} a_n^{(\rho)} C^{3/2}_n(2z-1)\right).
\end{align}
{In the above expression, the first term in the parentheses $1(=a_0^{(\rho)})$ corresponds to the normalisation \eqref{eq:rho_DA_norm}.
The expansion is related to the deviation from the perturbative DA $\phi_\rho(z)=6z(1-z)$ and the
leading coefficient $a_2^{(\rho)}$ provides information about the width of the DA.}

Through the light-cone QCD sum-rules, the $\rho$-meson DA $\phi_\rho$ plays an important role to describe the
processes such as exclusive semileptonic decays of the B-mesons ($B \to \rho l \bar \nu$) \cite{Ball:1997rj,Khodjamirian:1998ji,Ball:1998kk,Ball:2004rg}.
The resulting $B\to\rho$ transition form factors can be used to extract the CKM matrix element $|V_{ub}|$,
providing supplementary to the  process $B \to \pi l \bar \nu$ \cite{delAmoSanchez:2010af, Sibidanov:2013rkk} information
about $|V_{ub}|$.
{In the extraction of $|V_{ub}|$ from $B\to\rho$ transition form factors  the coefficient $a_2^{(\rho)}$ provide a nontrivial and prominent contribution, for detailed discussion, see \cite{Ball:2004rg}.}

Various non-perturbative method have been applied to study of the $\rho$-meson DA:  the QCD sum-rule
\cite{Chernyak:1981jd,Ball:1996tb,Ball:1998sk,Bakulev:1998pf,Ball:2004rg,Pimikov:2013usa,Fu:2016yzx},
 lattice simulation of the QCD \cite{Boyle:2008nj,Arthur:2010xf,Braun:2016wnx}, and other model approaches
\cite{Ji:1992yf,Choi:2007yu,Forshaw:2010py,Forshaw:2012im,Choi:2013mda,Gao:2014bca,Polyakov:1998ze,Shuryak:2019zhv}.
In Table~\ref{table:a2rho} we
collected various model predictions for $a_2^{(\rho)}$. One sees that the majority of the predictions corresponds to positive
value of the second Gegenbauer coefficient.

\begin{table}[ht]
\begin{tabular}{>{\centering}p{5cm}>{\centering}p{5cm}>{\centering}p{3cm}>{\centering}p{3cm}}
\hline\hline
 			References	&Approach		& $a_2^{\rho}$			& Scale \tabularnewline
\hline
%SR
Chernyak \textit{et al.} \cite{Chernyak:1981jd}	& SR &  $0.17 $   & $ \mu^2=1.5~\mathrm{GeV}^2$ \tabularnewline
Ball and Braun \cite{Ball:1996tb}	& SR &  $0.18 \pm 0.10 $   & $ \mu=1~\mathrm{GeV}$ \tabularnewline
Ball \textit{et al.} \cite{Ball:1998sk}	& SR &  $0.18 \pm 0.10 $   & $\mu=1~\mathrm{GeV}$ \tabularnewline
Bakulev and Mikhailov \cite{Bakulev:1998pf} & SR &  $0.079(20) $   & $\mu=1~\mathrm{GeV}$ \tabularnewline
Pimikov \textit{et al.} \cite{Pimikov:2013usa} & SR &  $0.047(61) $   & $ \mu=1~\mathrm{GeV}$ \tabularnewline
Stefanis and Pimikov \cite{Stefanis:2015qha} & SR &  $0.017(24) $   & $ \mu=2~\mathrm{GeV}$ \tabularnewline
Fu \textit{et al.} \cite{Fu:2016yzx} & SR &  $0.119(82) $   & $ \mu=1~\mathrm{GeV}$ \tabularnewline
%Lattice
 Boyle \textit{et al.} \cite{Boyle:2008nj}	& Lattice &  $0.108(105)(35)$   & $\mu=2~\mathrm{GeV}$ \tabularnewline
 Arthur \textit{et al.} \cite{Arthur:2010xf}	& Lattice &  $0.204(29)(58)$   & $\mu=2~\mathrm{GeV}$ \tabularnewline
 Braun \textit{et al.} \cite{Braun:2016wnx} & Lattice &  $0.132(27) $   & $\mu=2~\mathrm{GeV}$ \tabularnewline
%Model
Ji \textit{et al.} \cite{Ji:1992yf} & LCQM &  $-0.03 $   & $\mu=1~\mathrm{GeV}$ \tabularnewline
Choi and Ji \cite{Choi:2007yu} & LFQM &  $(+0.02,-0.02) $   & $ \mu=1~\mathrm{GeV}$ \tabularnewline
Qian et al. \cite{Qian:2020utg} & Light Front quantization &  $ 0.058 $   & $ \mu=1~\mathrm{GeV}$ \tabularnewline
Forshaw and Sandapen \cite{Forshaw:2010py} & Exp. Fit with LCWF &  $ (0.079,0.085,-0.053)  $  & $  \mu=1~\mathrm{GeV} $ \tabularnewline
Forshaw and Sandapen \cite{Forshaw:2012im} & Exp. Fit with AdS/QCD &  $0.082 $   & $ \mu=1~\mathrm{GeV}$ \tabularnewline
Choi and Ji \cite{Choi:2013mda} & LFQM &  $-0.02$   & $ \mu=1~\mathrm{GeV}$ \tabularnewline
Gao \textit{et al.} \cite{Gao:2014bca} & Dyson-Schwinger &  $0.088$   & $ \mu=1~\mathrm{GeV}$ \tabularnewline
Polyakov \cite{Polyakov:1998ze} &Instantons &$-0.14$    & $\mu\simeq 0.6~\mathrm{GeV}$ \tabularnewline
Shuryak \cite{Shuryak:2019zhv} & Instanton induced interaction & $\simeq -0.5$ & $\mu=?$ \tabularnewline
Clerbaux and Polyakov \cite{Clerbaux:2000hb} &Exp. Fit &$-0.1\pm 0.2$    & $\mu=4.5~\mathrm{GeV}$ \tabularnewline
\hline\hline
\end{tabular}
\caption{Values of $a_2^{(\rho)}$ obtained in various approaches.}
\label{table:a2rho}
\end{table}
In Ref.~\cite{Polyakov:1998ze} it was shown that  with help of the dispersion relations the Gegenbauer moments of the $\rho$-meson DA
  can be expressed in terms of the
two pion distribution amplitudes (2$\pi$DAs).
In particular, {$a_2^{(\rho)}$ is} related to the single pion observables thanks to the
crossing symmetry and the low-energy theorem. In Ref.~\cite{Polyakov:1998ze}, using the instanton approach,
 {the ratio of the $\rho$ and pion DA Gegenbauer moments $a_2^{(\rho)}/a_2^{(\pi)}\simeq -2.3$ was obtained, whereas the majority of the other works predicts the opposite sign for the ratio.}
 We argue here that the negative sign of the ratio $a_2^{(\rho)}/a_2^{(\pi)}$ is deeply rooted in chiral dynamics and general properties of
 quantum field theory such as unitarity, crossing and dispersion relations.

In this work, we first explain briefly how the $a_2^{(\rho)}$ is expressed in terms of the single pion observables owing
to low-energy theorems, crossing symmetry, and the dispersion relations.
{After that, we adopt the pion observables from a recent lattice calculation and global data analysis
to obtain the model independent relations for the ratio $a_2^{(\rho)}/a_2^{(\pi)}$.
Finally we present the results from the instanton model of the QCD vacuum.}

\end{section}

\begin{section}{Relation between $a_2^{(\rho)}$ moment and single pion observables }

The twist-2 chiral-even two-pion distribution amplitude(2$\pi$DA) is defined as follows \cite{Diehl:1998dk}:

\begin{align}\label{eq:2pida}
\Phi^{ab}(z,\zeta,W^2) = \left.  \frac{1}{4\pi} \int dx^- \exp(-i z P^+ x^- /2)
\;\langle \pi^a(p_1) \pi^b(p_2) | \bar{\psi}(x) \Slash{n} T \psi(0) | 0 \rangle \right|_{x^+=x_\perp = 0}.
\end{align}
Here we introduce the light-cone coordinate for a vector $v^\pm=n \cdot v = v^0\pm v^3$, with the
light-like vector $n^2=0$, represented as $n\to(1,0,0,1)$. $T$ is the isospin matrix and
in our case of interest, the isovector, $T=\tau^3/2$.
The 2$\pi$DA has three independent variables: $z$, the quark momentum fraction with respect to $P=p_1+p_2$,
$\zeta=p_1^+/P^+$, the longitudinal momentum distribution of two pions, and $W^2=P^2= (p_1+p_2)^2$, the invariant mass.
The isovector ($I=1$) part of the 2$\pi$DA can be projected out and
expanded in terms of the Gegenbauer polynomials $C^{\alpha}_n(x)$ in the following form
\begin{align}\label{eq:2pida_decomp}
\Phi^{I=1}(z,\zeta,W^2) = 6z(1-z)\sum^\infty_{n=0}\sum^{n+1}_{l=0}B_{nl}(W^2) C^{3/2}_n(2z-1)C^{1/2}_l(2\zeta-1).
\end{align}

In Ref.~\cite{Polyakov:1998ze} the dispersion relations for the generalised Gegenbauer moments $B_{nl}(W^2)$ were derived.
Moreover the solution of these dispersion relations for $W^2\leq 16 m_\pi^2$ was found:
\begin{align}\label{eq:Omnes}
B_{nl}(W^2)=B_{nl}(0) \exp \left[ \sum_{k=1}^{N-1} a_k^{(nl)} W^{2 k} +\frac{W^{2N}}{\pi} \int_{4 m_\pi^2}^\infty ds \frac{\delta_l^{I=1}(s)}{s^{N} (s-W^2-i 0)}\right].
\end{align}
Here $\delta_l^{I=1}(s)$ is the isospin one $\pi\pi$ scattering phase shift with the orbital momentum $l$, $N$ is the number of the subtractions in the dispersion relations,
and $a_k^{(nl)}$ are the corresponding  low-energy {subtraction} constants. For the $\rho$-meson channel ($I=1$ and $l=1$) it is known \cite{Guerrero:1997ku}
that the dispersion relations with two subtraction gives excellent description of the pion form factor for invariant mass till $W^2\simeq 2.5$~GeV$^2$. Therefore
we restrict ourselves to $N=2$ in Eq.~(\ref{eq:Omnes}) for the studies of generalised Gegenbauer moments at $W$ around
 the $\rho$-meson mass.

 The $\rho$-meson is a resonance in the $\pi\pi$ scattering amplitude  in the channel with $l=1, I=1$, the scattering phase shift $\delta_1^1(s)$
 in Eq.~(\ref{eq:Omnes}) crosses rapidly the value of $\pi/2$ near $s=m_\rho^2$.  {The scattering phase in the vicinity
 of the $\rho$ meson mass can be parametrised in the following form:

 \begin{align}
 \delta_1^1\left(s\sim m_\rho^2\right)={\rm arctg}\left(\frac{m_\rho \Gamma_\rho}{m_\rho^2-s}\right)+{\rm nonresonant\ contributions}
 \end{align}
 }
 {As it was shown in Ref.~\cite{Polyakov:1998ze}
 such behaviour of the scattering phase leads to the appearance of the pole in $B_{n1}(W^2)$ at $W=m_\rho-i \Gamma_\rho/2$, the residue
 in this pole corresponds to the Gegenbauer moment of the resonance DA, see detailed derivation in Ref.~\cite{Polyakov:1998ze}.
In this derivation it was important that the pole in $B_{n1}(W^2)$ coefficients for all $n$ appear solely
 from the dispersion integral in Eq.~(\ref{eq:Omnes}), therefore the phase of the residue in resonance pole is the same
 for all Gegenbauer moments, this common phase can be factored out in the Gegenbauer series (\ref{eq:rho_DA_G}) and the values
 of the Gegenbauer moments of the resonance DA are fixed by the values of $B_{n1}(0)$ at zero and by the subtraction constants  $a_k^{(nl)}$
 in Eq.~(\ref{eq:Omnes}). }
For example, the second Gegenbauer moment of the $\rho$-meson DA  can be obtained as \cite{Polyakov:1998ze}:

\begin{align}\label{eq:rhoM_1}
a_2^{(\rho)}= B_{21}(0) \exp(c_1^{(21)} m_\rho^2),
\end{align}
where $c_1^{(21)}=a_1^{(21)}-a_1^{(01)}$ is the subtraction constant. Its value is not know a priori, however it can be estimated in the
low-energy models or determined from the shape of $\pi\pi$ mass spectrum in hard exclusive processes, see detailed discussion and
fits to experimental data in Ref.~\cite{Clerbaux:2000hb}.

The expression (\ref{eq:rhoM_1}) can be further reduced to the single pion observables by the soft pion theorem and the crossing
symmetry \cite{Polyakov:1998ze}. {The soft pion theorem for 2$\pi$ DAs \cite{Polyakov:1998ze} relates
various Gegenbauer coefficients $B_{nl}(0)$
at zero invariant mass to the Gegenbauer moments of the pion DA\footnote{ This equation has a correction of order $\sim m_\pi^2$, there are no corrections enhanced by a chiral logarithms
\cite{Kivel:2002ia}. Numerically such corrections are of order $\sim 2 \%$. In what follows we shall consider consistently  the chiral limit.}:
\begin{align}
\label{eq:softpion}
\sum_{l=1}^{n+1} B_{nl}(0)=a_n^{(\pi)}.
\end{align}
Another important ingredient in our analysis is the crossing relations between $B_{n n+1}(0)$ and the $(n+1)$th Mellin moments
of parton distributions in the pion \cite{Polyakov:1998ze}:
\begin{align}
B_{n n+1}(0)=\frac{2(2 n+3)}{3(n+2)}\ M_{n+1}^{(\pi)} ,
\end{align}
with $M_{n+1}^{(\pi)} = \int^1_0 \;dx\;x^n (q_\pi(x)-\bar{q}_\pi(x))$.}
The above results allow us to relate the pion observables to the $\rho$-meson ones:
\begin{align}\label{eq:rhoM}
a_2^{(\rho)}= B_{21}(0) \exp(c_1^{(21)} m_\rho^2) = \left(a_2^{(\pi)} - \frac{7}{6}M_3^{(\pi)}\right)\exp(c_1^{(21)} m_\rho^2),
\end{align}
where $a_2^{(\pi)}$ is the second Gegenbauer moment of the $\pi$-meson light-cone DA and $M_3^{(\pi)}$ is the
{third} Mellin moment of the parton distribution functions in the pion,
\begin{align}\label{eq:M3}
M_3^{(\pi)} = \int^1_0 \;dx\;x^2 (q_\pi(x)-\bar{q}_\pi(x)).
\end{align}
{We note that for higher Gegenbauer moments $n\ge 4$ there are no such simple relations to single pion observables
as for the second moment in Eq.~(\ref{eq:rhoM}).
The reason is that for $n\ge 4$ the soft pion theorem (\ref{eq:softpion}) involves the Gegenbauer coefficients of resonances with the spins
$l=3,5,$etc., i.e. new unknown quantities.}
\\

In order to calculate the second Gegenbauer moment of the $\rho$-meson DA
with help of Eq.~(\ref{eq:rhoM}) we use
\begin{itemize}
\item
the results for $a_2^{(\pi)}$ from the lattice calculation,
\item
$M_3^{(\pi)}$ from global phenomenological analysis,
\item
the value of the subtraction constant $c_1^{(21)}$ from low-energy effective theory derived from instanton model
of QCD vacuum. We note that the precise value of the subtraction constant $c_1^{(21)}$ in (\ref{eq:rhoM}) does not influence the
sign of $a_2^{(\rho)}$.
\end{itemize}
The value of $a_2^{(\pi)}$ we take from the lattice simulation of Bali \textit{et al.} (RQCD) \cite{Bali:2019dqc} {at the next-to-leading order(NLO) \footnote{{The main result of Bali \textit{et al.} \cite{Bali:2019dqc} is presented at the next-to-next-to-leading order (NNLO) as
$a_2^{(\pi)}(\mu = 2 ~\mathrm{GeV}) = 0.101 \pm 0.024$. In the present work, we take their NLO result as the pion PDF we adopt \cite{Bali:2019dqc} is extracted at the NLO, see below. We are grateful to N. G. Stefanis for pointing out this to us.}}}:
\begin{align}\label{eq:bali_a2}
a_2^{(\pi)}(\mu = 2 ~\mathrm{GeV}) = 0.078 \pm 0.028 {\; (\mathrm{RQCD\ at\ NLO})}.
\end{align}
 The central value is smaller by a factor of $\sim 1/2$ compared to the central values of older lattice results
 \cite{Boyle:2008nj,Arthur:2010xf,DelDebbio:2002mq,Braun:2006dg}
 evaluated at the same renormalisation point or higher. We note that the lattice
 simulations \cite{Bali:2019dqc} are the most advanced in terms of the extrapolation to the chiral and continuum limits.
 For the $M_3^{(\pi)}$, we consider the results of the recent phenomenological analysis by I.~Novikov et al. (xFitter) \cite{Novikov:2020snp},
 which supersedes the old analysis by M.~Gluck et al. \cite{Gluck:1999xe}.
At the renormalisation scale $\mu = 2~\mathrm{GeV}$ they obtained {at the NLO}:
\begin{align}
M_3^{(\pi)}(\mu = 2 ~\mathrm{GeV}) = 0.114 \pm 0.020 \; (\mathrm{xFitter})\label{eq:M3_phen2}.
\end{align}
{The error bars in above equation is obtained by expressing $M_3^{(\pi)}$ in terms of the model parameters
of Ref.~\cite{Novikov:2020snp} and then computing the uncertainty of $M_3^{(\pi)}$ from error bars for the parameters given in Ref.~\cite{Novikov:2020snp}.}

In Table \ref{table:b21}, the ratios $B_{21}(0)/M_3^{(\pi)}$
and $a_2^{(\rho)}/a_2^{(\pi)}$ obtained with help of Eq.~(\ref{eq:rhoM}) are shown.
We give the results for ratios, instead of individual observables, as the former
stay stable under the scale evolution at the one-loop order.
Note that the positive factor $\exp(c_1^{(21)} m_\rho^2)$ should be multiplied to
$B_{21}(0)$ to obtain the $a_2^{(\rho)}$, as seen in Eq. \eqref{eq:rhoM}.
As the factor is not yet determined phenomenologically\footnote{The subtraction constant $c_1^{(21)}$
can be measured in hard exclusive two pion production, see discussion in Ref.~\cite{Clerbaux:2000hb}}, we use the model value calculated in this work,
$c_1^{(21)}\in [0.7, 0.9]$~GeV$^{-2}$. The detailed discussion of this constant will be given in the following section.
{Note that the smaller value of $a_2^{(\pi)}$ results in a cancellation for $B_{21}(0)$.}
Even though it is difficult at this point to conclude on the sign of $a_2^{(\rho)}$ decisively
 due to relatively large statistical errors in the studies, the results still strongly suggest that the $a_2^{(\rho)}$ 
 is negative\footnote{In what follows we shall implicitly assume that $a_2^{(\pi)}\ge 0$, for a negative
$a_2^{(\pi)}$ the second Gegenbauer moment of $\rho$-meson DA is obviously negative, see Eq.~(\ref{eq:rhoM}) }.

\begin{table}[ht]
\begin{tabular}{>{\centering}p{2cm}>{\centering}p{3.5cm}}
\hline\hline
 									& RQCD/xFitter \tabularnewline
\hline
 $ B_{21}(0)/M_3^{(\pi)} $	  & $-0.48 \pm 0.27$ \tabularnewline
 $ a_2^{(\rho)}/a_2^{(\pi)} $  & $( -1.15 \pm 0.86)(1.0\pm 0.1)$ \tabularnewline
\hline\hline
\end{tabular}
\caption{Phenomenologically determined  ratios $B_{21}(0)/M_3^{(\pi)}$
 and  $a_2^{(\rho)}/a_2^{(\pi)}$ at QCD normalisation point $\mu=2$~GeV. The ratios are stable under scale evolution at the one-loop order.
For calculation of the ratio $ a_2^{(\rho)}/a_2^{(\pi)} $ the subtraction constant is
taken in the range  $c_1^{(21)}\in [0.7, 0.9] \;\mathrm{GeV}^{-2}$, this range is calculated within the instanton model in the present work.
The uncertainty related to the variation of $c_1^{(21)}$ is shown in the last bracket in the bottom row (note that this bracket is always positive).  }
\label{table:b21}
\end{table}

\end{section}

\begin{section}{$a_2^{(\rho)}$ from instantons}
The low-energy effective action of quarks interacting with (pseudo)Goldstone bosons
derived from the theory of the instanton vacuum \cite{Diakonov:1983hh,Diakonov:1985eg}
has the following form in the Euclidean space:
\begin{align}\label{eq:action}
S_{\mathbf{eff}} =  \int d^4x\ \bar \Psi(x) \left[i \Slash{\partial} + i \sqrt{M(i\partial)} U^{\gamma_5}(x) \sqrt{M(i\partial)}\right] \Psi(x).
\end{align}
$U^{\gamma_5}$ is the non-linear chiral $SU(2)_f$ field defined by
\begin{align}\label{eq:ugamma5}
U^{\gamma_5}(x) = U(x) \frac{1+\gamma^5}{2} + U^\dagger(x) \frac{1-\gamma^5}{2},
\end{align}
with \begin{align}\label{eq:ufield}
U(x) = \exp\left[\frac{i}{F_\pi} \pi^a(x)  \tau^a\right].
\end{align}
The momentum dependence of the dynamical quark mass arises from the Fourier transform of the quark zero-mode and
has the following representation in the momentum space \cite{Diakonov:1985eg}:
\begin{align}
M(k) &= M_0 F^2(k), \label{eq:DQM}\\
F(k) &= 2t \left.\left[ I_0(t)K_1(t) - I_1(t)K_0(t)-\frac{1}{t}I_1(t)K_1(t)\right]\right|_{t=\frac{k\rho}{2}}.\label{eq:qFF}
\end{align}
In the above expression $K$ and $I$ are modified Bessel functions and $\rho$ is the average instanton size.
The zero-momentum quark mass $M_0=M(k=0)$ generated by the spontaneous breakdown of the chiral symmetry
is calculated by using the gap equation:
\begin{align}\label{eq:gap}
\frac{N}{V} = 4N_c \int \frac{d^4k}{(2\pi)^4}\; \frac{M^2(k)}{k^2+M^2(k)},
\end{align}
where $N$ is the average instanton number and $V$ is the 4-Euclidean volume.
In the above expression, the average instanton density $N/V$ can be re-expressed as $N/V=1/R^4=(\rho^4/R^4)/\rho^4$,
where $R$ is the average instanton inter-distance.
Hence, $M_0$ is obtained with given average instanton packing fraction and average instanton size.
Typically the set of values $\rho/R\approx 1/3$ and $R\approx1$~fm was obtained from
Feynman variational principle for QCD partition function\cite{Diakonov:1983hh,Diakonov:1995qy}.
In principle, the values of $\rho$ and $R$ are related to the $\Lambda_{\mathrm{QCD}}$ and
can be matched to perturbative QCD to provide the model renormalisation point,  see Ref. \cite{Diakonov:1995qy}.

The effective action (\ref{eq:action}) can be  used to compute many low-energy observables for the pions.
For example, the pion decay constant
 which is well established experimentally $F_\pi\simeq 93$~MeV ($F_\pi\simeq 88$~MeV in the chiral limit \cite{Gasser:1983yg}),
can be computed as the loop integral in 4D-Euclidean momentum space
which has the following parametrical dependence:
\begin{align}\label{eq:Fpi_para}
F_\pi^2 = \frac{C}{\rho^2}~ \left(\frac{\rho}{R}\right)^4  \ln \left( \frac{R^2}{\rho^2} \right),
\end{align}
with the constant $C\sim 1$. We see that in the theory of the instanton vacuum the pion decay constant
has a natural suppression by the small instanton packing fraction, hence the instanton mechanism of the spontaneous
chiral symmetry breaking explains the ``accidental" smallness of the pion decay constant.
With the typical choice of $\rho/R\approx 1/3$ with $R=1$~fm, we obtain $F_\pi\sim 100$~MeV.
The best values of the instanton vacuum parameters were determined with help of Feynman variational principle
in Refs.~\cite{Diakonov:1983hh,Diakonov:1985eg}, the values are $\rho/R\simeq 1/3$ and $\rho\simeq 0.33$~fm.
In this paper we vary these parameters in the range which corresponds to pion decay constant (in the chiral limit)
$F_\pi=88\pm 15$~MeV.

The calculations of the pion and two-pion DAs were pioneered in Ref.~\cite{Petrov:1998kg} and Ref.~\cite{Polyakov:1998td} respectively.
Main finding in these papers was that the momentum dependence of the quark mass is very important to determine the shape of
(two)pion DAs. Using the simplifying assumption about the momentum dependence of the quark mass the first estimates
of the DAs were performed. The technique was refined in Refs.~\cite{Praszalowicz:2001wy,Praszalowicz:2001pi,RuizArriola:2002bp,Praszalowicz:2003pr},
and we refer to these papers for details of calculations of pion DAs.
{In all these previous works a pole-type quark-mass momentum dependence was used
to simplify the calculation and to obtain some analytic results.
In this work, we will use directly \eqref{eq:qFF} to explorer the instanton parameter dependences of the observables
as it follows from the theory of the instanton vacuum.}

In the limit $\rho\to 0$ with fixed $\rho/R \ll 1$ one obtains that:
\begin{align}
a_2^{(\pi)} = \frac{7}{18}, \quad
M_3^{(\pi)} = \frac{1}{3}, \quad [{\rm limit}\ \rho\to 0] .\label{eq:limit_M3}
\end{align}
Therefore, owing to the soft pion theorem and crossing relations, we have in this limit:
\begin{align}
B_{21}(0)=0 \quad [{\rm limit}\ \rho\to 0].
\end{align}
For small $W^2$ one can also easily obtain in the limit of small $\rho$ \cite{Polyakov:1998ze}:
\begin{align}
B_{21}(W^2)= - \frac{7 N_c}{1440 \pi^2 F_\pi^2} W^2+O\left(W^4\right) \quad [{\rm limit}\ \rho\to 0].
\end{align}
The result indicates that $B_{21}(W^2)$ is getting negative with increasing of the two-pion invariant mass.

The above limiting results correspond to flat pion DA and to flat quark distribution function in the pion.
Also these limiting results implies that $a_2^{(\rho)}\to 0$ for $\rho\to 0$, see Eq. \eqref{eq:qFF}.
Furthermore we made an important observation that beyond the limit of zero instanton size $B_{21}(0)$ is always negative.
Therefore, independently of the details of the dynamics (e.g. the shape of the form factor $F(k)$) the second
Gegenbauer moment for $\rho$-meson DA is always negative.

In Fig.~\ref{fig:b21a2M3} we show the results of our calculations of the ratio $B_{21}(0)/M_3^{(\pi)}$ for various
values of instanton parameters $\rho$ and $\rho/R$. We also show on this figure by the vertical shaded band
the range of instanton parameters which leads to reasonable values of $F_\pi=88\pm 15$~MeV, about 20\% around the
phenomenological value of the decay constant in the chiral limit { for $\rho=0.33$ fm.}
\begin{figure}[ht!]
\begin{center}
	\includegraphics[width=12cm]{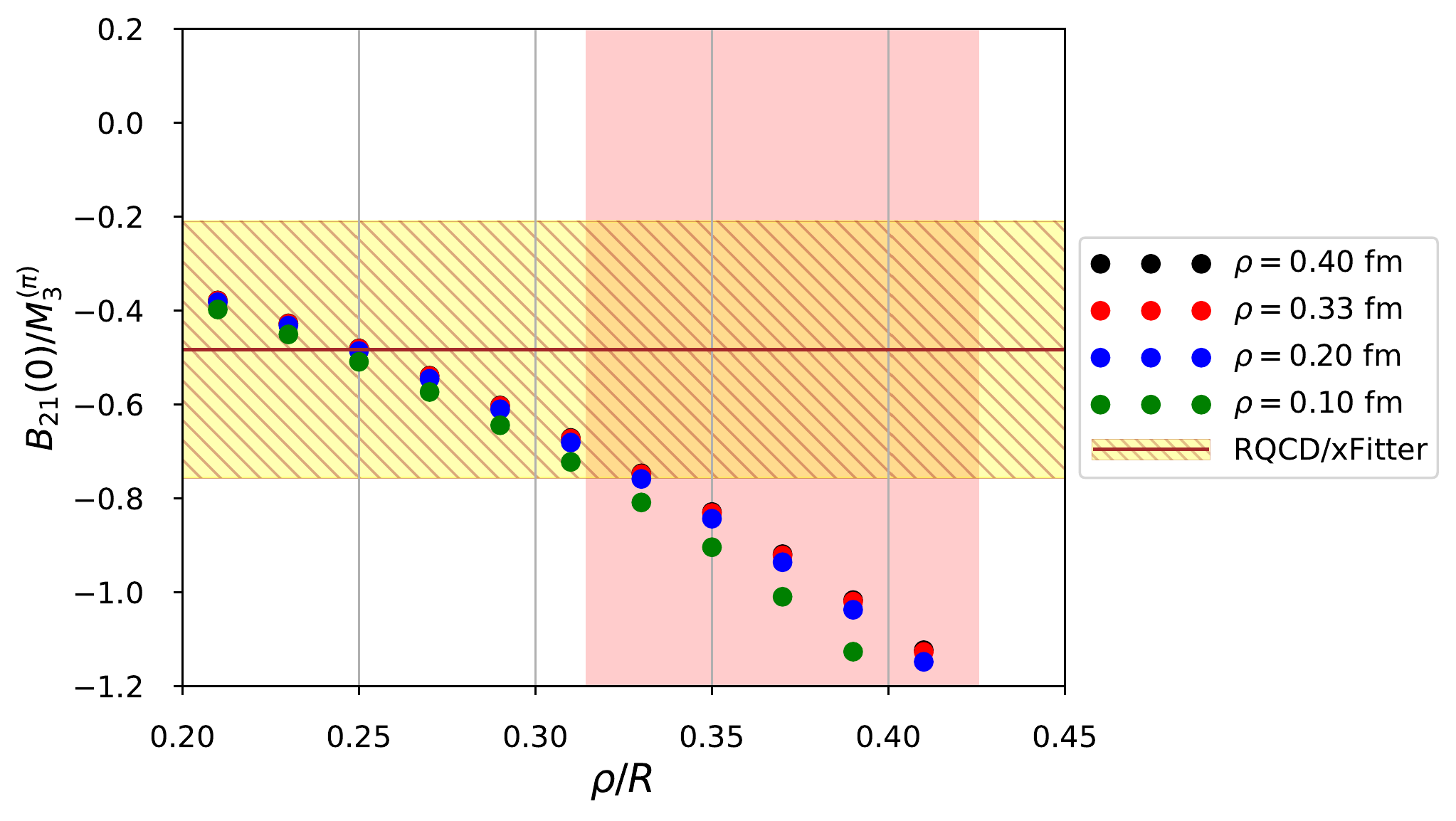}
\caption{ Comparison of the ratio $B_{21}(0)/M_3^{(\pi)}$,
 calculated for various values of the parameters $\rho/R$ and $\rho$ to the phenomenological values from Table \ref{table:b21} (horizontal
 band).
The vertical band shows the range of $\rho/R$ resulting
 to $F_\pi =88\pm 15$~MeV for $\rho=0.33$~fm.
 } \label{fig:b21a2M3}
\end{center}
\end{figure}
From Fig.~~\ref{fig:b21a2M3} we clearly see that the ratio $B_{21}(0)/M_3^{(\pi)}$ in the instanton model is definitely negative
and is compatible (within 2$\sigma$) with the phenomenological analysis from previous section. The latter is shown by the
horizontal shaded 1$\sigma$ band.

In order to {obtain} the second Gegenbauer moment of the $\rho$-meson DA {from} Eq.~(\ref{eq:rhoM}), we
need {to calculate} the subtraction constant $c_1^{(21)}$. In Fig.~\ref{fig:c21} the result of the calculation of the
dimensionless combination $c_1^{(21)} F_\pi^2$ is shown for various values of instanton vacuum parameters $\rho$ and $\rho/R$.
\begin{figure}[ht!]
\begin{center}
	\includegraphics[width=9cm]{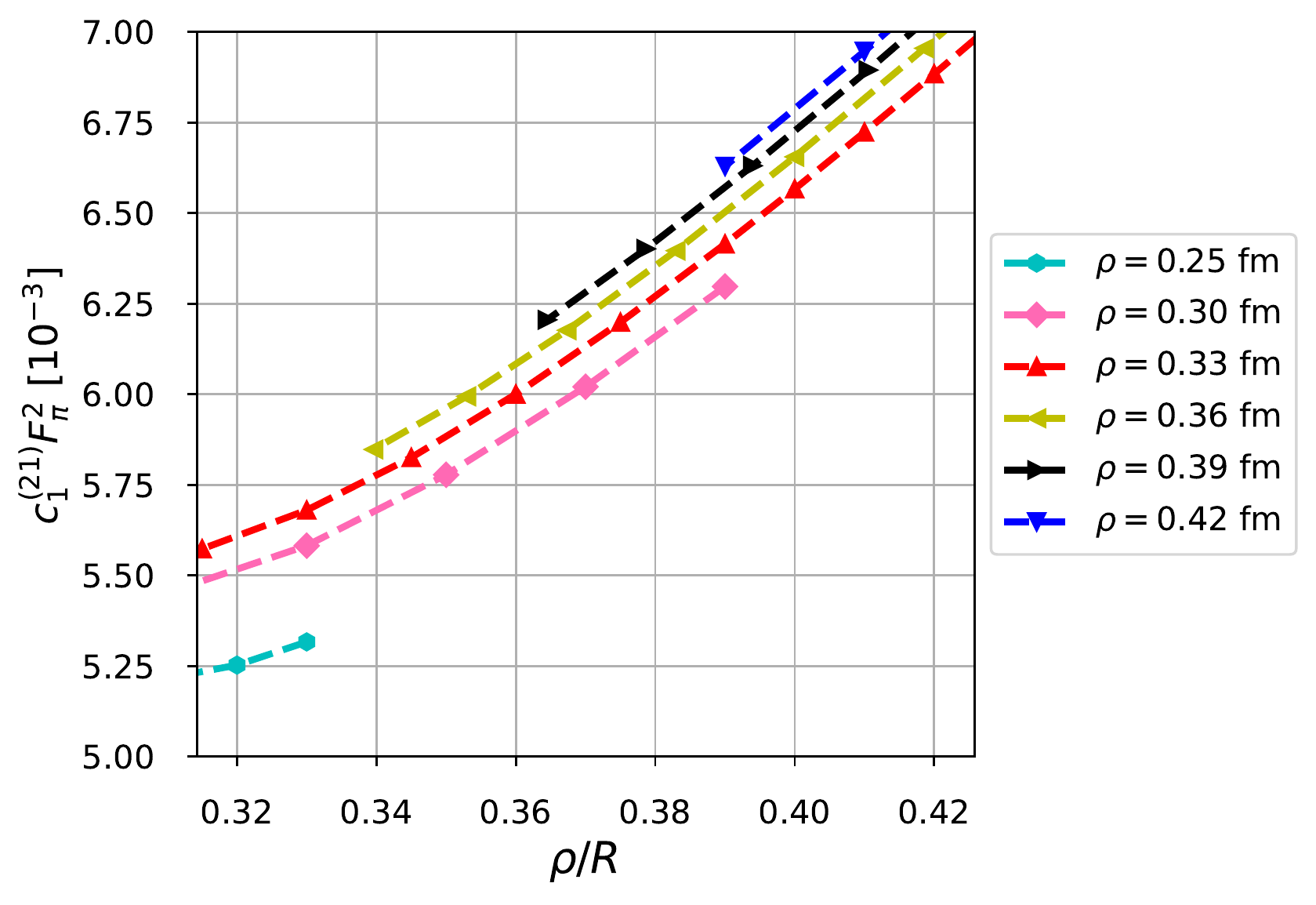}
\caption{ Dimensionless combination of the subtraction constant $c_1^{(21)}$
and $F_\pi$ calculated for various $\rho$ and $\rho/R$.  The range for latter corresponds to $F_\pi =88\pm 15$~MeV. } \label{fig:c21}
\end{center}
\end{figure}
We see that the subtraction constant has rather strong dependence on $\rho/R$, this prevents us
from precise determination of this constant. Therefore for the phenomenological analysis presented in the previous
section we assumed generously that $c_1^{(21)}\in [0.7, 0.9]$~GeV$^{-2}$ which follows from results presented in
Fig.~\ref{fig:c21}. We note that the sharp dependence of
the subtraction constant on $\rho/R$ can be used for the phenomenological determination of this instanton parameter
from a measurement of $c_1^{(21)}$.

Eventually, combining the calculations of $B_{21}(0)$ and the subtraction
constant, we present in Fig. \ref{fig:a2rhodiva2pi} the ratio $a_2^{(\rho)}/M_3^{(\pi)}$ at various values of the
instanton parameters.
\begin{figure}[ht!]
\begin{center}
\includegraphics[width=10cm]{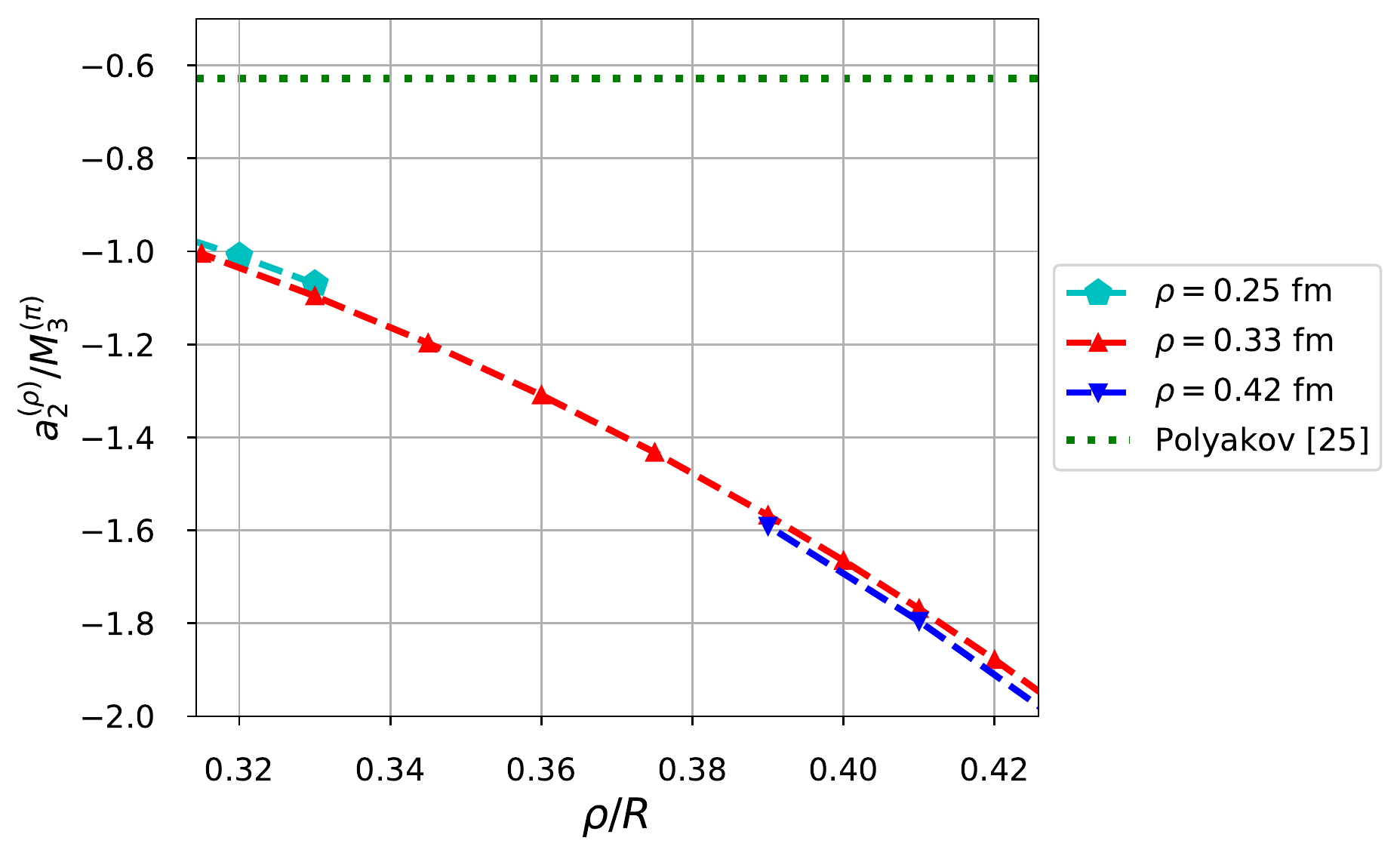}
\caption{The ratio
$a_2^{(\rho)}/M_3^{(\pi)}$  as a function of $\rho/R$
 is plotted for various $\rho$ values.
 The green dotted horizontal line indicates the result
$a_2^{(\rho)}/M_3^{(\pi)}=-0.63$  from Ref. \cite{Polyakov:1998ze}.
Presented range of $\rho/R$ corresponds to $F_\pi=88\pm 15$~MeV for $\rho=0.33$~fm.
} \label{fig:a2rhodiva2pi}
\end{center}
\end{figure}
The value of the ratio $a_2^{(\rho)}/M_3^{(\pi)}$
has a rather wide range, roughly from $-2$ to $-1$, depending on the given values of the packing fraction $\rho/R$
and the average instanton size $\rho$. Important observation is that the ratio $a_2^{(\rho)}/M_3^{(\pi)}$, and hence $a_2^{(\rho)}$,
is {\it always negative} in the instanton model. This is at variance with the majority of the results
for this quantity  in various models, see Table~\ref{table:a2rho}.
It is also very interesting that the recent calculation of the $\rho$-meson DA in an approach which emphasises the role of topologically (instanton) induced
quark interactions \cite{Shuryak:2019zhv} also predicts the negative $a_2^{(\rho)}$.

\end{section}

\begin{section}{Summary}
Main findings of the present work are:

\begin{itemize}
	\item
	Using the soft pion theorem, crossing, and the dispersion relations for two pion distribution
amplitude ($2\pi$DA) we performed phenomenological analysis for the ratio of the second Gegenbauer moments of the pion and the $\rho$ meson DAs  with the result at $\mu=2~{\rm GeV}$\footnote{Note that the ratio $a_2^{(\rho)}/a_2^{(\pi)}$ is renormalisation scale independent at one-loop order.}:

\begin{align}\label{eq:mainres1}
a_2^{(\rho)}/a_2^{(\pi)}= ( -1.15 \pm 0.86)(1.0\pm 0.1).
\end{align}
As the input for our analysis we used the value of $a_2^{(\pi)}$ from the most advanced to date lattice simulation in Ref.~\cite{Bali:2019dqc} and
the most recent phenomenological analysis of the pion PDFs \cite{Novikov:2020snp}. Eventually the yet experimentally unknown  subtraction constant
for the dispersion relations is calculated here using the effective low-energy theory derived from the instanton theory of the QCD vacuum.
{The range for the subtraction constant obtained within the model leads to the uncertainty of the numerical estimate (\ref{eq:mainres1}),
it is reflected  in the last brackets of Eq.~(\ref{eq:mainres1}). Note that this bracket is always positive -- uncertainty
in the subtraction constant does not influence the sign in  Eq.~(\ref{eq:mainres1})}
	\item
	We computed the ratio $a_2^{(\rho)}/M_3^{(\pi)}$ in the instanton model of the QCD vacuum:
\begin{align}\label{eq:mainres2}
a_2^{(\rho)}/M_3^{(\pi)} \in [-1, -2].
\end{align}
Here the range of the ratio reflects rather generous range of variation for the parameters of the instanton vacuum.
The model calculations demonstrate that $a_2^{(\rho)}$ is negatively defined, this is in qualitative disagreement, especially what concerns the sign, with the majority
of calculations of $a_2^{(\rho)}$ in the literature (see summary of results in Table~\ref{table:a2rho}). It is important to find the reasons for this qualitative discrepancy. 
\end{itemize}
From these results we may conjecture that the topologically non-trivial field configurations in the
QCD vacuum (instantons) lead to qualitatively different shapes of the pion and the $\rho$-meson distribution
amplitudes. Similar picture was obtained, although from different perspective, by E.~Shuryak in Ref.~ \cite{Shuryak:2019zhv}.

\end{section}

\section*{\normalsize \bf Acknowledgements}
We are grateful to Hyun-Chul Kim, Sergey Mikhailov, and Nico Stefanis for interesting discussions.
This work is supported by the DFG through the Sino-German CRC 110
 ``Symmetries and the Emergence of Structure in QCD".
 MVP  thanks the organisers of LCDA20 workshop at Mainz Institute for Theoretical Physics
for invitation, and  for discussions there which had triggered these notes.

\end{document}